\documentclass[aps,twocolumn,english,prl,superscriptaddress]{revtex4-1}

\usepackage{graphicx}
\usepackage{dcolumn}
\usepackage{bm}
\usepackage[T1]{fontenc}
\usepackage[latin9]{inputenc}
\usepackage{textcomp}
\usepackage{graphicx}
\usepackage{amssymb}
\usepackage{units}
\usepackage{xcolor}
\usepackage{amsmath}

\begin{document}

\title{Phonon mediated conversion of exciton-polaritons Rabi oscillation into THz radiation}
\author{Katharina Rojan}
\affiliation{Université Grenoble-Alpes, CNRS, Laboratoire de Physique et Modélisation des Milieux Condensés, F-38000 Grenoble, France}
\affiliation{Theoretische Physik, Universit\"at des Saarlandes, D-66123 Saarbr\"ucken, Germany}
\author{Yoan L\'eger}
\affiliation{FOTON Laboratory, CNRS, INSA, F-35708 Rennes, France}
\author{Giovanna Morigi}
\affiliation{Theoretische Physik, Universit\"at des Saarlandes, D-66123 Saarbr\"ucken, Germany}
\author{ Maxime Richard}
\affiliation{Université Grenoble-Alpes, CNRS, Institut Néel, F-38000 Grenoble, France}
\author{Anna Minguzzi}
\affiliation{Université Grenoble-Alpes, CNRS, Laboratoire de Physique et Modélisation des Milieux Condensés, F-38000 Grenoble, France}


\begin{abstract}
Semiconductor microcavities in the strong-coupling regime exhibit an energy scale in the THz frequency range, which is fixed by the Rabi splitting between the upper and lower exciton-polariton states. While this range can be tuned by several orders of magnitude using different excitonic medium, the transition between both polaritonic states is dipole forbidden. In this work we show that in Cadmium Telluride microcavities, the Rabi-oscillation driven THz radiation is actually active without the need for any change in the microcavity design. This feature results from the unique resonance condition which is achieved between the Rabi splitting and the phonon-polariton states, and leads to a giant enhancement of the second order nonlinearity.
\end{abstract}

\date{\today}

\maketitle

A room temperature compact solid-state THz source is a highly desirable piece of equipment, needed in many domains like communication, health or security \cite{federici2005}. Different semiconductor based techniques already exist, based on cascade laser structures, optical nonlinearity, or photomixing in electronic structures, but  so far they all require advanced cooling techniques \cite{Tonouchi2007,lewis2014}.

Although coincidental, it turns out that semiconductor microcavities in the strong-coupling regime \cite{Weisbuch1992} exhibit the right energy scale for THz electromagnetic radiation. Indeed the normal mode - (Rabi) splitting, i.e.~the energy splitting between the upper and lower exciton-polariton states resulting from the strong-coupling regime, typically ranges from 3.5 meV (i.e.~$0.85\,$THz) in GaAs microcavities, to hundreds of meV (i.e.~tens of THz) in large bandgap semiconductors systems. And yet, such a transition cannot be used directly to generate or absorb THz photons as it features no dipole moment.

To circumvent this problem, several strategies have been proposed. In a pioneering experiment, an in-plane static electric field was used to hybridize excitonic states with different parities (i.e.~$s$-like and $p$-like) resulting in a nonzero dipole moment between the upper and lower polariton branches \cite{Hokomoto1999}. More recently, taking advantage of the bosonic nature of polaritons, it has been proposed to optically excite $p$-like excitons by two-photons absorption, and to achieve bright THz emission by stimulated relaxation toward the lower polariton branch \cite{Kavokin2012,Kaliteevski2014}. However, due to ultrafast relaxation of the $p$-excitons, this mechanisms has remained elusive so far \cite{Huppert2014}. Other promising ideas have been put forward like e.g. intersubband polaritons microcavities involving doped asymmetrical quantum wells \cite{DeLiberato2013}, or microcavities embedding a $\chi^{(2)}$ active materials like [111]-oriented GaAs \cite{Barachati2015}.

\begin{figure*}[t]
\centering
\includegraphics[width=0.85\textwidth]{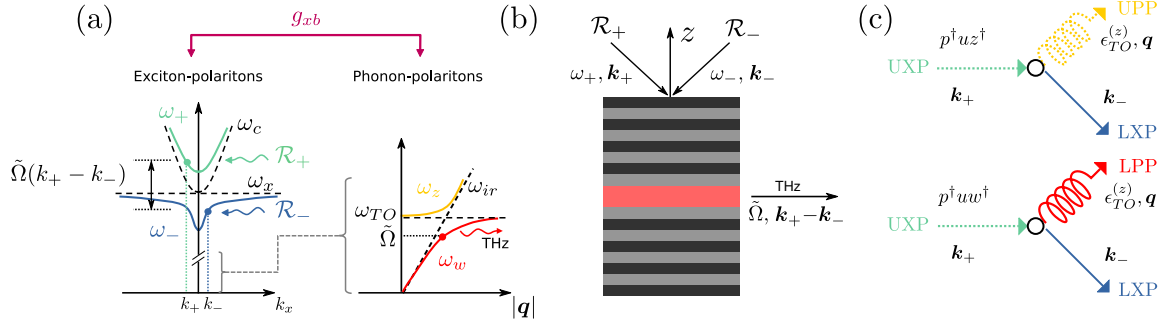}
\caption{(a) Sketch of the conversion mechanism: left,  upper (green, $\omega_+$) and lower (blue, $\omega_-$) exciton-polariton dispersion branches along the  $x$ axis; right, upper (yellow, $\omega_z$) and lower (red, $\omega_W$) phonon-polariton branches vs wavevector $|\bm q|$. Dots in exciton-polariton branches indicate an exemple of laser-pumped states with intensity $\mathcal{R}_+$ and $\mathcal{R}_-$ respectively, with frequency difference  $\tilde{\Omega}$. The dot in the lower phonon-polariton branch show the state emitting THz radiation at frequency $\tilde{\Omega}/2\pi$. (b) Geometric representation of the pump scheme: the excitation laser beams impinges the cavity surface each with an angle such that $\sin(\theta_\pm)=ck_\pm/\omega_\pm$. The generated phonon-polaritons propagate in the cavity plane with $\bm q=\bm k_{+}-\bm k_{-}$ and decay at the interface into THz radiation. (c) Elementary interaction mechanisms of this scheme : An upper exciton-polariton (UXP) is annihilated in order to create a lower exciton-polariton (LXP) plus a phonon-polariton (upper or lower depending on $\tilde{\Omega}$).}
\label{fig1}
\end{figure*}

The common point in these proposals is to modify the microcavity structure, either by applying an external field, or by engineering the material, in order to build up a $\chi^{(2)}$ optical nonlinearity resonant with the polariton states, and to use it for frequency difference generation. In this work, we show that a threefold-resonant condition can be reached when the Rabi splitting is tuned at resonance with a mechanical vibration mode related to the lattice unit-cell deformation, i.e.~the transverse-optical (TO) phonon. This mechanism does not require any externally applied field, or any new materials.

The most extensively used material to fabricate microcavities in the strong-coupling regime is Gallium Arsenide \cite{Carusotto2013}. In this material, this resonance condition is hard to meet as the typical achievable Rabi splitting is in general much smaller than the TO phonon energy ($\hbar\omega_{TO}=33.3\,$meV). A CdTe-based microcavity will be considered in this letter as such microcavities exhibit a near perfect match between the typical microcavity Rabi splitting \cite{Kasprzak2008} and the TO phonon energy $\hbar\omega_{TO}=18\,$meV, with the additional benefit of a much better thermal stability tested up to $T=200\,$K \cite{saba2001}.

Our proposal relies on a cascade of three different mechanisms already present in current state-of-the art microcavities, namely: (i) the strong-coupling regime between optical cavity photons and excitons, (ii) the strong-coupling regime between THz photons and TO phonons (i.e.~the so-called phonon-polariton states which are intrinsic to most semiconductor materials, including the ones cited in this letter) \cite{hopfield1965}, and (iii) the deformation-potential mediated interaction between the exciton hole and the TO phonons. The latter provides the effective $\chi^{(2)}$ nonlinearity enabling the conversion of laser-pumped upper and lower exciton-polariton field into THz photons. We focus  on TO phonons because,  although longitudinal optical (LO) phonons are much better coupled to excitons via Fr\"olich mechanism,
they  are not coupled to THz photons and are thus useless in our scheme. 
Interestingly, (iii) is formally the same interaction mechanism that appears in the context of optomechanics, except that photons are replaced by excitons and that both oscillators benefits from a continuum spectrum of phononic excitations. Frequency conversion schemes in the microwave domain have indeed been achieved by purely optomechanical means using high-Q mechanical resonators \cite{Bochmann2013,Andrews2014,Bagci2014}. In our case, mechanical vibrations are also at the heart of the mechanism as TO phonons correspond to the mechanical deformation of the few Angstrom-sized crystal unit cell.


A detailed physical picture of our conversion scheme is sketched in Fig.\ref{fig1}(a): by optically exciting simultaneously  both  upper and lower polaritons, an excitonic density is created in the microcavity active region (a slab of bulk semiconductor), which is time modulated at the Rabi frequency. This population beat induces mechanical vibration within the active region via deformation-potential interaction. In our scheme, this modulation frequency matches that of a TO-phonon-polariton state, so that the earlier excites the latter. Finally, the resulting phonon-polaritons propagate in the structure and decay radiatively into THz photons upon reaching an interface or by interacting with a structured dielectric environment. 
Accounting for this cascade of interactions, the system's Hamiltonian consists of four terms~:
\begin{align}\label{Htot}
\mathcal{\hat H}=\mathcal{\hat H}_{\textrm{exc-cavity}}+\mathcal{\hat H}_{\textrm{phon-THz}}+\mathcal{\hat H}_{\textrm{exc-phon}}+\mathcal{\hat H}_{\textrm{laser}},
\end{align}
where
\begin{align}
& \mathcal{\hat H}_{\textrm{exc-cavity}}=\sum_{\alpha=1,2}\sum_{\;\;\bm{k}_{\parallel}}\hbar\omega_{{x}} \hat c^{\dagger}_{\alpha,\bm{k}_{\parallel}}\hat c^{ }_{\alpha,\bm{k}_{\parallel}} \nonumber \\
& +\hbar\omega_{c}(\bm{k}_{\parallel},\delta)\hat a^{\dagger}_{\alpha,\bm{k}_{\parallel}}\hat a_{\alpha,\bm{k}_{\parallel}}+\hbar\Omega(\hat a^{\dagger}_{\alpha,\bm{k}_{\parallel}}\hat c_{\alpha,\bm{k}_{\parallel}}+h.c.).
\end{align}
describes the interaction between optical cavity photons (creation operator $ \hat a^{\dagger}_{\alpha,\bm{k}_{\parallel}}$ for an in-plane momentum $\bm{k}_{\parallel}$ and a polarization state $\alpha$) and a bright excitonic state (creation operator $\hat c^{\dagger}_{\alpha,\bm{k}_{\parallel}}$) in the strong-coupling regime; $2\hbar\Omega$ is the Rabi splitting between the upper and lower exciton-polaritons states for zero detuning between cavity and exciton. The energy of the excitonic transition does not depend on the wave vector and is given by $\hbar\omega_x=1630$ meV in CdTe. We choose $2\hbar\Omega=12\,$meV $<\hbar\omega_{TO}$, so that both phonon-polaritons branches (lower and upper) can be used for the conversion mechanism. Such a Rabi splitting is easily achievable in CdTe microcavities \cite{richard2005}.


Neglecting the dependency on the polarization \cite{Panzarini1999}, the cavity mode has  dispersion
\begin{equation}
 \omega_{c}(\bm{k}_{\parallel},\delta)=\omega_{c}^0+\frac{\hbar|\bm{k}_{\parallel}|^2}{2m_{\text{eff}}}+\delta,
\label{omega-c}
\end{equation}
where $\omega_{c}^0= ck_{z,0}/n_\text{vis}$ with $k_{z,0}$ fixed by the cavity length and $n_\text{vis}\simeq 2.5$ is the cavity medium refractive index in the visible range; $m_{\text{eff}}\simeq 10^{-5}$ $m_e$  is the effective mass of the cavity photons, with $m_e$ the free electron mass; $\delta$ is the cavity detuning, modifiable through the microcavity intentional wedged shape \cite{Deveaud2007}. From now on we take  $\omega_{c}^0=\omega_x$.

The interaction between TO phonons and THz photons in  Eq.~\eqref{Htot}  has the form
\begin{align}
\mathcal{\hat H}_{\textrm{phon-THz}}&=\sum_{\bm{q}}\hbar\omega_{{TO}}\hat b^{\dagger}_{\bm{q}}\hat b_{\bm{q}}+\hbar\omega_{{ir}}(\bm{q})\hat l^{\dagger}_{\bm{q}}\hat l_{\bm{q}}\nonumber\\
&+\hbar\Omega_{ir}(\hat b^{\dagger}_{\bm{q}}\hat l_{\bm{q}}+\hat l^{\dagger}_{\bm{q}}\hat b_{\bm{q}}),
\end{align}
where $\hat b^{\dagger}_{\bm{q}}$ creates a TO phonon with a three-dimensional momentum $\bm{q}$ and frequency $\omega_{TO}$, $\hat l^{\dagger}_{\bm{q}}$ creates a THz photon with momentum $\bm{q}$ and  dispersion
\begin{equation}
 \omega_{ir}(\bm{q})= (c/n_\text{ir})\sqrt{|\bm{q}_{\parallel}|^2+q_z^2},
\label{omega-ir}
\end{equation}
where $n_\text{ir}=3.57$ is the cavity medium refractive index in the THz range \cite{Schall2001}. Phonons and THz photons are in strong coupling, giving rise to phonon-polaritons, with $2\hbar\Omega_{ir}=11.15$ meV being the phonon-polariton Rabi splitting \cite{Schall2001}.

The cornerstone of the conversion mechanism is the weak interaction between bright excitons and TO phonon, described by the optomechanical-like Hamiltonian \cite{appendix}
\begin{align}
\label{H_xb}
 \mathcal{\hat H}_{\rm{exc-phon}}& = -\sum_{\substack{\bm{q},\bm{k},\bm{k'}}} \hbar g_{xb} \left(\hat b^{\dagger}_{-\bm{q},TO}+\hat b_{\bm{q},TO}\right)\delta_{\bm{q},\bm{k}-\bm{k'}}\nonumber\\
&\times \left(\hat c^{\dagger}_{\bm{k},1}\hat c_{\bm{k'},2}+\hat c^{\dagger}_{\bm{k},2}\hat c_{\bm{k'},1}\right),
\end{align}
where $\bm{k},\bm{k}'$ and $\bm{q}$ are  three-dimensional wavevectors and $\alpha=1,2$ the exciton linear polarizations oriented along the crystalline axes $x=$[010] and $y=$[001]. Notice that the exciton-phonon coupling is then diagonal in the basis where the linear polarizations are oriented at $\pm$ 45$^\circ$. { This feature
 is in agreement with the results of Raman scattering theory, which involves the same interaction vertex \cite{Cardona}.
The exciton-TO phonon interaction in zincblende crystals occurs via the deformation potential. Owing to symmetries, the only non-vanishing matrix element are those coupling cross-polarized heavy-hole (\textit{hh}) and light-hole (\textit{lh}) exciton states \cite{Cantarero1989,Ganguly1967}, via the deformation potential in the direction perpendicular to the polarization plane  (see \cite{appendix}). Notice that in order  to involve both excitonic state with a comparable weight  it is important to use  a bulk CdTe active layer inside the microcavity, while quantum wells would not be suitable since  the \textit{hh} and \textit{lh} exciton states are split by the confinement. The deformation potential has as non-zero matrix element,  in the basis of cubic harmonics, eg  $\langle X|V_{DP}^{(z)}|Y \rangle$,  where  $V_{DP}^{(z)}$ is the $z$-component of the deformation potential vector (see \cite{appendix}). Interestingly, this result agrees with the symmetry properties of the nonlinear susceptibility $\chi^{(2)}$ of zincblende crystals in the spectral region we are interested in, whose only non-zero elements are $d_{14}$, $d_{25}$, $d_{36}$ following the standard nomenclature \cite{Boyd}. In particular, one may have  $P_z=d_{36} E_x E_y$, ie a phonon polarization along $z$ induced by applied fields in $x$ and $y$ directions thanks to the off-diagonal terms of the nonlinear susceptibility tensor. This is our case, since there is a preferential direction fixed by the cavity axis that fixes the  incoming and outgoing photons to be polarized in the $(x,y)$ plane. Correspondingly, the emitted phonons participating to the coupling to excitons are propagating in the cavity plane, i.e.~$\mathbf{q}=(\mathbf{q_\parallel},q_z=0)$. The exciton-phonon coupling $\hbar g_{xb}$ that we obtain  (see \cite{appendix}) reads :
\begin{align}
 \hbar g_{xb}=q_h \sqrt{\frac{\hbar M}{2\rho V\mu\omega_{TO}a^2}}\frac{d_0}{2},
\end{align}
where $d_0$ is the optical deformation potential \cite{BirPikuspaper,BirPikus}, $a$ is the length of the elementary cell of the lattice, $\rho$ is the mass density of the semiconductor, $V$ the volume of the slab, $\mu$ and $M$ are the reduced and total mass of the two ions of the elementary cell of the crystal. The prefactor $q_h$ corresponds to the overlap between the orbitals describing the relative motion of electron and hole before and after the interaction of the exciton and the TO phonon. For a CdTe slab of size $V=18\unit{\mu m^3}$ using   $d_0=30$ eV \cite{Poetz1981} we estimate $\hbar g_{xb}=0.2\ \mu$eV \cite{appendix}.

\begin{figure}[tpb]
 \begin{center}
\includegraphics[width=0.45\textwidth]{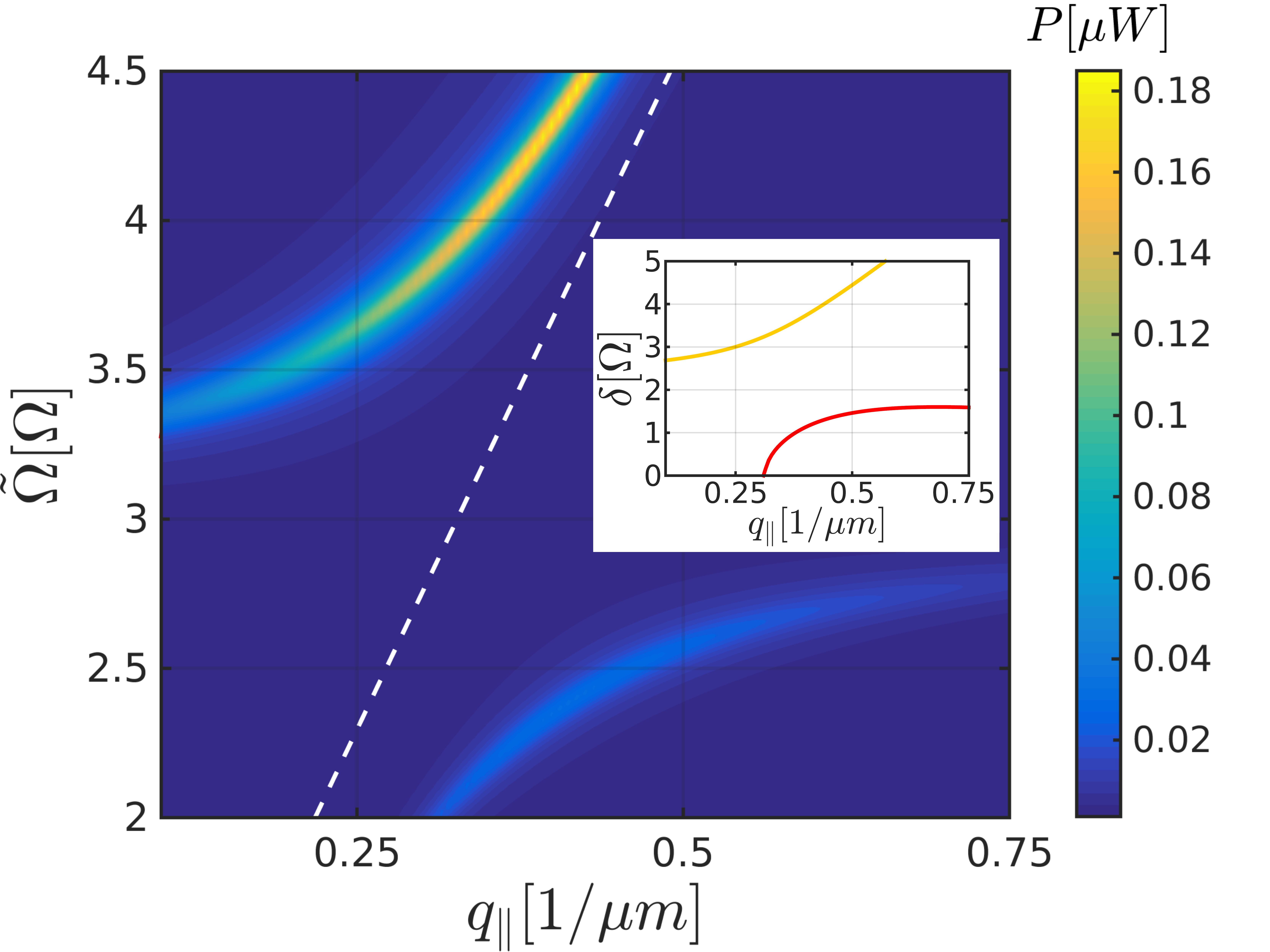}
 \end{center}
\caption{Emission power of THz photons as function of their frequency $\tilde\Omega$ (in units of $\Omega$) and their in-plane wave vector $q_{\parallel}=|\bm{q}_{\parallel}|$ (in units of $\mu$m$^{-1}$) for input power  $P_{in}^{\pm}=\hbar\omega_{\pm}\frac{2|\mathcal{R}_{\pm}|^2}{C_{\bm k}^2\gamma_p}=0.9$ mW, with $\mathcal{R}_{\pm}=1\Omega$, $C_{\bm k}^2=0.5$ and resonant pumping $\omega_-=\omega_p$ and $\omega_+=\omega_u$. The decay rates are $\kappa=0.1\unit{THz},\gamma_b=0.35\unit{THz},\gamma_l=0.6\unit{THz}$,  $\gamma_w=\gamma_z\simeq0.6\unit{THz}$ \cite{Schall2001} taking for simplicity a value which is momentum independent. The typical decay rate of the exciton $\gamma_x$ is neglected since  $\gamma_x \sim 0.01\unit{THz}$. The white dashed line corresponds to the light cone $\omega_{ir}(q_{\parallel})= (c/n_{\text{ir}})|q_{\parallel}|$. Inset: cavity detuning $\delta$ in units of $\Omega$ as function of  $q_{\parallel}$ for $\tilde\Omega=\omega_w$ (red curve) and $\tilde\Omega=\omega_z$ (yellow curve).\label{Fig3}}
\end{figure}

Being orders of magnitude weaker as compared to $\Omega$ and $\Omega_{ir}$, the interaction amplitude $g_{xb}$ will be conveniently treated as a perturbation. Following Hopfield's transformation, we can rewrite $\mathcal{\hat H}_{\textrm{exc-cavity}}$ in the basis of upper and lower exciton-polariton \cite{Hopfield1958}~:
\begin{equation}\label{ex_pol_trafo}
\begin{pmatrix}
 \hat p_{\alpha,\bm{k}}\\
 \hat u_{\alpha,\bm{k}}\\
\end{pmatrix}=
\begin{pmatrix}
 -C_{\bm{k}} &X_{\bm{k}}\\
 X_{\bm{k}}&C_{\bm{k}}\\
\end{pmatrix}
\begin{pmatrix}
 \hat a_{\alpha,\bm{k}}\\
 \hat c_{\alpha,\bm{k}}\\
\end{pmatrix},
\end{equation}
where $\hat p^{\dagger}_{\alpha,\bm{k}}$ and $\hat u^{\dagger}_{\alpha,\bm{k}}$ are the bosonic creation operator of lower and upper exciton-polariton states, with in-plane momentum $\bm{k}$ and energy $\hbar \omega_{p,u}(\bm{k})$, where $p,u$ stands for lower/upper polariton respectively. $C_{\bm{k}}$ and $X_{\bm{k}}$ are the usual Hopfield coefficients describing the photonic and excitonic fraction of these exciton-polariton states \cite{Deveaud2007}. In the same way, we apply another Hopfield transformation \cite{hopfield1965} to the TO phonons/THz photons subspace, that reads~:
\begin{equation}
\label{ex_pol_traf2}
\begin{pmatrix}
\hat w_{\bm{q}}\\
\hat z_{\bm{q}}\\
\end{pmatrix}=
\begin{pmatrix}
 -N_{\bm{q}}&T_{\bm{q}}\\
 T_{\bm{q}}&N_{\bm{q}}\\
\end{pmatrix}
\begin{pmatrix}
\hat b_{\bm{q}}\\
\hat l_{\bm{q}}\\
\end{pmatrix},
\end{equation}
where $\hat w^{\dagger}_{\bm{q}}$ and $\hat z^{\dagger}_{\bm{q}}$ are the bosonic creation operator of lower and upper phonon-polariton states respectively, with energy $\hbar \omega_w$ and $\hbar \omega_z$. $N_{\bm{q}}$ and $T_{\bm{q}}$ are the Hopfield coefficients describing the TO phonons and THz photons fraction respectively of these phonon-polariton states. In our total Hamiltonian, $g_{xb}$ thus couples the exciton-polaritons and phonon-polaritons subspace; interestingly, since it is very weak as compared to the other coupling amplitudes $g_{xb}\ll \Omega, \Omega_{ir}$, we can safely assume that the polaritonic states, whether excitons or phonons, remain the proper eigenstates of the system.

The last term in Eq.(\ref{Htot}) describes the optical excitation that pumps the exciton-polariton states resonantly. We consider a two-pump scheme, as depicted in Fig.\ref{fig1}(b), which excites both the  lower polariton branch of a given polarization, e.g. $\alpha=1$, and the upper polariton branch of opposite polarization, i.e.~$\alpha'=2$ \cite{commentpump}. The pumps  have  amplitudes  $\hbar \mathcal{R}_{\pm}$, wave-vector $\bm{k}_\pm$,  and frequency  $\omega_{\pm}$. The corresponding  Hamiltonian reads
\begin{align}
\label{H_laser}
\mathcal{\hat H}_{\textrm{laser}}=\hbar\mathcal{R}_{-}e^{-i\omega_{-}t}\hat p^{\dagger}_{1,\bm{k}_-}+\hbar\mathcal{R}_{+}e^{-i\omega_{+}t}\hat u^{\dagger}_{2,\bm{k}_+}+h.c.
\end{align}

In the following,  we  will focus on the  two exciton-polariton states which are macroscopically populated by the laser pump, ie $\hat p_1$ and $\hat u_2$. Among the various exciton-phonon interaction terms in $\mathcal{\hat H}_{\textrm{exc-phon}}$ generated by the Hopfield transformations (\ref{ex_pol_trafo}) and (\ref{ex_pol_traf2}) we keep those which describe the frequency-conversion process we are interested in, namely  $\hat p_1^{\dagger}\hat u_2\hat w^{\dagger}$, $\hat p_1^{\dagger}\hat u_2\hat z^{\dagger}$. These processes are summarized into the diagrams of Fig.\ref{fig1}(c).


In order to determine the input-output characteristics of the system in the steady-state, accounting for thermal noise contribution, we derive the corresponding Heisenberg-Langevin equations
\begin{widetext}
\begin{align}
&\dot {\hat p}_{1,\bm{k}_{-}}=-i\omega_{p}(\bm{k}_{-})\hat p_{1,\bm{k}_{-}}+ig_{xb}X_{\bm{k}_-}C_{\bm{k}_+}\hat u_{2,\bm{k}_{+}}\left(-N_{-\bm{q}}\hat w_{-\bm{q}}^{\dagger}+T_{-\bm{q}}\hat{z}_{-\bm{q}}^{\dagger}\right)-i\mathcal{R}_{-}e^{-i\omega_{-}t}-\gamma_{p}\hat p_{1,\bm{k}_-}+\sqrt{2\gamma_{p}}\hat p_{1,\textrm{in}}(t),  \label{HB_p}\\
&\dot {\hat u}_{2,\bm{k}_{+}}=-i\omega_{u}(\bm{k}_{+})\hat u_{2,\bm{k}_+}+ig_{xb}C_{\bm{k}_+}X_{\bm{k}_-}\hat p_{1,\bm{k}_-}\left(-N_{\bm{q}}\hat w_{\bm{q}}+T_{\bm{q}}\hat z_{\bm{q}}\right)-i\mathcal{R}_{{+}}e^{-i\omega_{+}t}
-\gamma_{u}\hat u_{2,\bm{k}_+}+\sqrt{2\gamma_{u}}\hat u_{2,\textrm{in}}(t),   \label{HB_u}\\
&\dot {\hat w}_{\bm{q}}=-i\omega_{w}(\bm{q})\hat w_{\bm{q}}-ig_{xb}N_{\bm{q}}X_{\bm {k}_-}C_{\bm{k}_+}\hat p_{1,\bm{k}_-}^{\dagger}\hat u_{2,\bm{k}_+}-\gamma_{w}\hat w_{\bm{q}}+\sqrt{2\gamma_w}\hat w_{\textrm{in}}(t), \label{HB_w}\\
&\dot {\hat z}_{\bm{q}}=-i\omega_{z}(\bm{q})\hat z_{\bm{q}}+ig_{xb}T_{\bm{q}}X_{\bm{k}_-}C_{\bm{k}_+}\hat p_{1,\bm{k}_-}^{\dagger}\hat u_{2,\bm{k}_+}-\gamma_{z}\hat z_{\bm{q}}+\sqrt{2\gamma_z}\hat z_{\textrm{in}}(t) \label{HB_z},
\end{align}
\end{widetext}
where the condition $\bm{q}=\bm{k}_+-\bm{k}_-$ is assumed all through  the Eqs. \eqref{HB_p}-\eqref{HB_z}, $\gamma_{p,u,w,z}$ are the decay rates of exciton-polariton and photon-polaritons, and $\hat \zeta_{\rm in}=\hat p_{\textrm{in}},\hat u_{\textrm{in}}, \hat w_{\textrm{in}},\hat z_{\textrm{in}}$ are input noise fields, with $\langle \hat\zeta_{\rm in}(t)\rangle=0$ and $\langle [\hat\zeta_{\rm in}(t),\hat\zeta_{\rm in}^\dagger(t')]\rangle=\delta(t-t')$ \cite{Collett:1985}.
The decay rates of upper and lower exciton-polariton can be estimated using  excitonic $\gamma_x$ and photonic $\kappa$ decay rates according to  $\gamma_{u}=\gamma_xC_{\bm{k}_+}^2+\kappa X_{\bm{k}_+}^2$ and $\gamma_{p}=\gamma_xX_{\bm{k}_-}^2+\kappa C_{\bm{k}_-}^2$. Correspondingly we use $ \gamma_{w}=\gamma_bN_{\bm{q}}^2+\gamma_l T_{\bm{q}}^2,$ and $\gamma_{z}=\gamma_bT_{\bm{q}}^2+\gamma_l N_{\bm{q}}^2$ for the decay rates of lower and upper phonon polariton. At room temperature, the only non-negligible noise is the one originating from phonons and THz photons, populated according to  $\bar n_b=\frac{1}{e^{\hbar\omega_{TO}/k_BT}-1}$ and $\bar n_l=\frac{1}{e^{\hbar\omega_{ir}/k_BT}-1}$, respectively.
Since $g_{xb}\ll \Omega, \Omega_{ir}$, we can decouple the  Heisenberg-Langevin equations \eqref{HB_p}-\eqref{HB_z}. We first obtain the expectation value of $\hat p_{1,\bm{k}_-}^{\dagger}\hat u_{2,\bm{k}_+}$ for times longer than the characteristic time scale of the Rabi oscillations ($t\gg T_R=1/\tilde\Omega$), corresponding to $n_0(t)\equiv\langle \hat p_{1,\bm{k}_-}^{\dagger}(t)\hat u_{2,\bm{k}_+}(t)\rangle_{t\gg T_R}$, from  Eqs.~\eqref{HB_p} and \eqref{HB_u} assuming $g_{xb}\simeq0$ and neglecting correlations. Pumping both exciton polariton branches leads to an effective pumping of the phonon-polariton branches at frequency  $\tilde\Omega=\omega_{+}-\omega_{-}$. This frequency-difference generation scheme follows from the specific nonlinearity of the exciton-phonon coupling Eq.(\ref{H_xb}).
Inserting the solution for $n_0(t)$ in Eqs.~\eqref{HB_w} and \eqref{HB_z}, we find the number of phonon-polaritons in lower and upper mode  $N_w=\langle w_{\bm{q}}^{\dagger}(t)w_{\bm{q}}(t) \rangle$ and  $N_z=\langle z_{\bm{q}}^{\dagger}(t)z_{\bm{q}}(t)\rangle$ at large times $t\gg T_R$:
\begin{eqnarray}
\label{N_l}
 N_w&=\frac{|\tilde{\mathcal{R}}|^2N^2(\bm{q})}{\gamma_w^2+(\tilde\Omega-\omega_w(\bm{q}))^2}+\frac{\gamma_b}{\gamma_w}N^2(\bm{q})\bar n_b(\omega_{TO})  \nonumber\\&+\frac{\gamma_l}{\gamma_w}T^2(\bm{q})\bar n_l(\omega_{ir}) 
\end{eqnarray}
\begin{eqnarray}
 N_z&=\frac{|\tilde{\mathcal{R}}|^2T^2(\bm{q})}{\gamma_z^2+(\tilde\Omega-\omega_z(\bm{q}))^2}+\frac{\gamma_b}{\gamma_z}T^2(\bm{q})\bar n_b(\omega_{TO})\nonumber\\&+\frac{\gamma_l}{\gamma_z}N^2(\bm{q})\bar n_l(\omega_{ir}),
\end{eqnarray}
with effective pump strength
\begin{equation}
\label{eta-tilde}
 \tilde{\mathcal{R}}=\frac{g_{xb}X_{\bm{k}_-}C_{\bm{k}_+}\mathcal{R}_{-}^*\mathcal{R}_{+}}{(-i\gamma_{p}-(\omega_p-\omega_-))(i\gamma_{u}-(\omega_u-\omega_+))}.
\end{equation}
Equations \eqref{N_l}-\eqref{eta-tilde} show that our frequency-difference generation scheme is based on a triple resonance, yielding a giant effective $\chi^{(2)}=1035\,$ pm/V for the intracavity fields (see \cite{appendix}). Owing to the triple resonance conditions, this value is ten times larger than that found in other materials used for difference frequency generation of THz light \cite{tochitsky2007,petersen2011}. Note that for fields considered outside the cavity $\chi^{(2)}$ is even much larger, i.e.~by a typical factor $\propto X_{\bm{k}_-}C_{\bm{k}_+}Q$, where $Q$ is the bare cavity quality factor. $Q\simeq 2000$ is routinely achieved in CdTe microcavities \cite{richard2005}.

The emission power of THz photons, i.e.~the photonic fraction of lower and upper phonon-polaritons, is given by  $P=T^2 N_w\gamma_w\hbar\omega_w+N^2 N_z\gamma_z\hbar\omega_z$. Figure \ref{Fig3} shows the tunability and emission power as a function of the frequency $\tilde\Omega$ and the in-plane wave vector for the case of  resonant pumping. The white dashed line corresponds to the light cone Eq.(\ref{omega-ir}).
The THz frequency $\tilde\Omega$ depends both on the wavevector $q_{\parallel}$ and on the detuning $\delta$ between cavity and exciton.
One main advantage of the proposed scheme is that the frequency $\tilde \Omega$ is easily tunable in actual experiments: its lowest value is fixed by the Rabi splitting $\Omega$, and it can be adjusted by varying the detuning $\delta$ of the  cavity photon energy $\hbar\omega_c$   in Eq.(\ref{omega-c})  using the microcavity intentional wedged shape \cite{Deveaud2007}. For $\tilde\Omega$ matching the frequency of upper or lower phonon polariton branch, $\delta$ depends on the wave vector $q_{\parallel}$ as depicted in the inset of figure \ref{Fig3}.

As explained above, the THz emission takes place in the plane of the microcavity with a wavevector $\mathbf{q_\parallel}$. In order to collect easily this emission the simplest strategy is to etch or deposit microstructures on the cavity surface, such as a grating or so-called ``bullseyes'' \cite{luca2015}, in order to scatter the THz photons back into the forward direction. The typical pitch of such structures has to be comparable and larger than the THz wavelength, i.e $\sim 50\,\mu$m, which poses no technological difficulty.


In conclusion, we have presented a frequency down conversion scheme to convert optical photons in THz photons in a CdTe semiconductor microcavity. The scheme is based on an engineered $\chi^{(2)}$ optical nonlinearity coming from the weak interaction between bright excitons and TO phonons and doesn't involve any change in the microcavity material nor an externally applied field. For future applications the case of large bandgap material is promising in two respects: (i) the strong-coupling regime is stable at room temperature, with obvious benefits for the future realization of an actual polaritonic THz devices and (ii) the TO phonons energy ($\hbar\omega_{TO}=(66.5;69.5)\,$ meV in GaN and $\hbar\omega_{TO}=(47.7;51.7)\,$ meV in ZnO \cite{note1}) can be well matched by the Rabi splitting in a state-of-the-art microcavity \cite{Li2013,Trichet2011,Christmann2008}. Due to their wurtzite crystalline structure additional calculations are needed to determine exactly the value of the bright exciton TO phonon coupling strength, but symmetry considerations suggest that our mechanism should be applicable as well.

\acknowledgements
The authors acknowledge discussions with Jo\"el Cibert. This work is supported by French state funds ANR-10-LABX-51-01 (Labex LANEF du Programme d'Investissements d'Avenir), and ANR-16-CE30-0021 (QFL), the French-German University (UFA/DFH), the German Research Foundation, and the German Ministery of Education and Research (BMBF "Qu.com").


\begin{thebibliography}{39}
\bibitem [{com()}]{commentpump}%
  \BibitemOpen
  \href@noop {} {}\bibinfo {note} {In alternative, one may use the same
  polarization direction for the two pumps, oriented at 45 degrees with respect
  to the $x,y$ plane.}\BibitemShut {Stop}%
\end{thebibliography}
\end{document}